\documentclass{article}

\usepackage{microtype}
\usepackage{graphicx}
\usepackage{float}
\usepackage{xcolor}
\DeclareRobustCommand{\headwayreview}[1]{\textcolor{black}{#1}}
\usepackage{booktabs}
\usepackage{multirow}
\usepackage{amsmath}
\usepackage{amssymb}
\usepackage{xspace}
\usepackage{enumitem}
\usepackage{tikz}
\usetikzlibrary{patterns}
\usepackage{hyperref}
\renewcommand{\equationautorefname}{Eq.}

\newcommand{\theHalgorithm}{\arabic{algorithm}}

\usepackage[accepted]{mlsys2025}
\hypersetup{pdfsubject={}}
\renewcommand{\icmlruler}[1]{}
\newcommand{\headwayprofile}[2]{%
  \texorpdfstring{\href{https://openreview.net/profile?id=~#2}{\headwayreview{#1}}}{#1}}

\newcommand{\sys}{\textsc{Headway}\xspace}

\newcommand{\stagebubble}{stage bubble\xspace}
\newcommand{\mbbubble}{microbatch bubble\xspace}
\newcommand{\membubble}{memory bubble\xspace}
\newcommand{\pdbubble}{P/D handoff bubble\xspace}
\newcommand{\pending}{\emph{pending}\xspace}
\definecolor{findingborder}{HTML}{3B7757}
\definecolor{findingbg}{HTML}{EEF7F0}
\usepackage[framemethod=default]{mdframed}
\newcommand{\finding}[2]{%
  \begin{mdframed}[
    linecolor=findingborder,backgroundcolor=findingbg,
    linewidth=0.45pt,roundcorner=0pt,
    leftmargin=0pt,rightmargin=0pt,
    innerleftmargin=5pt,innerrightmargin=5pt,
    innertopmargin=5pt,innerbottommargin=5pt,
    skipabove=0.35em,skipbelow=0.35em,
    splittopskip=5pt,splitbottomskip=5pt,nobreak=false,everyline=false]
    \let\findingoriginaltextbf\textbf
    \renewcommand{\textbf}[1]{\textcolor{findingborder}{\findingoriginaltextbf{##1}}}%
    \noindent\textcolor{findingborder}{\textbf{Finding #1.}} #2
  \end{mdframed}}

\usepackage{etoolbox} \usepackage{needspace} \newlength{\savedtextfloatsep} \newlength{\savedfloatsep} \newlength{\savedintextsep}       \makeatletter \let\savedmakecaption\@makecaption \patchcmd{\@makecaption}{\vskip 10pt}{\vskip 4pt}{}{} \pretocmd{\appendix}{   \setlength{\textfloatsep}{\savedtextfloatsep}   \setlength{\floatsep}{\savedfloatsep}   \setlength{\intextsep}{\savedintextsep}   \let\@makecaption\savedmakecaption}{}{} \makeatother \mlsystitlerunning{Mind the Gap: Minimizing Bubbles in Pipeline-Parallel LLM Serving}

\begin{document}
\raggedbottom

\twocolumn[
\mlsystitle{Mind the Gap: Minimizing Bubbles in \mbox{Pipeline-Parallel LLM Serving}}

\begin{mlsysauthorlist}
\ifdefined\isaccepted
  \color{black}
  \mlsyssetsymbol{equal}{*}
  \mlsysauthor{\headwayprofile{Youhe Jiang}{YOUHE_JIANG1}}{meta,equal}
  \mlsysauthor{\headwayprofile{Yi Xu}{Yi_Xu64}}{meta,equal}
  \mlsysauthor{\headwayprofile{Haoran Wu}{Haoran_Wu16}}{cambridge}
  \mlsysauthor{\headwayprofile{Zhan Shu}{Zhan_Shu4}}{meta}
  \mlsysauthor{\headwayprofile{Ehsan K. Ardestani}{Ehsan_K._Ardestani1}}{meta}
  \mlsysauthor{\headwayprofile{Adnan Aziz}{Adnan_Aziz1}}{meta}
  \par\smallskip
  \textsuperscript{1}Meta\qquad\textsuperscript{2}University of Cambridge
\else
  \textbf{Anonymous Authors}
\fi
\end{mlsysauthorlist}

\mlsyskeywords{LLM serving, pipeline parallelism, speculative decoding, chunked prefill, scheduling}

\ifdefined\isaccepted\vskip 0.12in\else\vskip 0.3in\fi

\begin{abstract}
Whether accelerators are connected by PCIe or NVLink, serving a model that
outgrows a tightly coupled accelerator group requires pipeline parallelism (PP),
yet pipeline bubbles keep PP from turning its communication advantage into
throughput. We target four types of bubbles, across stages, microbatches,
memory, and the prefill/decode (P/D) handoff, and show that the levers serving
engines already expose (layer partitioning, prefill chunk size, speculative
draft length, request admission, and prefill microbatching) each move more than
one bubble and interact, so tuning them in isolation leaves throughput
unclaimed. We present \sys, which co-optimizes these levers: it
selects the layer partition offline \headwayreview{because repartitioning is computationally costly. It}
jointly schedules chunk size and draft length online under TTFT and TPOT guards
and coordinates request admission and
prefill microbatching in P/D-disaggregated serving. On replayed coding-agent
sessions with two models, \sys improves token throughput over
state-of-the-art pipeline-parallel serving systems by 17.5\% on average (up to
32.6\%) in co-located serving and by 19.2\% in disaggregated serving, while
reducing mean time between output tokens by 17.6\% at comparable
time-to-first-token latency.
\end{abstract}
]

\ifdefined\isaccepted
  \begingroup
  \renewcommand{\thefootnote}{\fnsymbol{footnote}}
  \footnotetext[1]{Equal contribution.}
  \endgroup
\fi

\section{Introduction}
\label{sec:intro}

State of the art Large Language Models (LLMs) now have trillions of parameters
and growing faster than the high-bandwidth domains that serve them. Kimi~K2 has 1.04T parameters~\citep{kimiteam2025kimik2}, and its successor Kimi~K3, released a year later, has 2.8T~\citep{moonshot2026kimik3}. Even in its native 4-bit MXFP4 format, K3's weights occupy roughly 1.5\,TB, more than the HBM of an eight-GPU H100 (640\,GB), H200 (1.13\,TB), or B200 (1.44\,TB) node before any KV cache is allocated~\citep{nvidia2023h100,nvidia2024h200,nvidia2024dgxb200}.
Rack-scale NVLink systems such as GB300 extend this domain to
72 GPUs~\citep{nvidia2025gb200nvl72,nvidia2025gb300nvl72}. If parameter counts keep growing at the K2-to-K3 rate, frontier models will
outgrow even a 72-GPU domain within a couple of model generations. Beyond the scale-up
domain, the remaining links (PCIe within a server, InfiniBand or Ethernet across
nodes and racks) are too slow for tensor parallelism~\cite{zhang2025td,hwang2026revisiting}, so PCIe- and NVLink-connected systems need pipeline
parallelism to serve the latest models and remain useful throughout their
five- to six-year depreciation period~\cite{microsoft2022tenk,alphabet2023tenk,amazon2024tenk}.

Pipeline parallelism (PP) assigns consecutive layers to stages and exchanges activations at stage boundaries, reducing communication volume relative to TP~\cite{li2023alpaserve}.
This makes PP \headwayreview{beneficial} in two settings:
(\textbf{\underline{i}}) Within a node, PP's lower communication demand suits PCIe-connected accelerators such as NVIDIA RTX PRO 6000 GPUs and Rebellions ATOM NPUs~\cite{nvidia2025rtxpro,rebellions2024atom}.
(\textbf{\underline{ii}}) At larger model scales, BF16 \headwayreview{configurations} of models such as Muse Spark Open with 816B parameters can require multiple GPU nodes~\cite{meta2026musespark,huggingface2026musespark}.
PP reduces communication over slower inter-node links~\cite{narayanan2021efficient}.

However, pipeline bubbles can prevent PP from translating its communication advantage into high serving throughput~\cite{agrawal2024taming,zhang2025td}.
We distinguish stage, microbatch, and memory bubbles, as well as handoff bubbles in prefill/decode (P/D) disaggregated serving (\S\ref{sec:motiv:types}).
These bubbles reflect both the execution schedule and the serving system's configuration.

Reducing these bubbles requires adjusting work distribution and memory use.
There are several levers within pipeline-parallel serving engines~\cite{kwon2023efficient,zheng2024sglang}: pipeline layer partitioning~\cite{li2023alpaserve,jiang2024hexgen}; prefill chunk size~\cite{agrawal2024taming}; speculative decoding draft length~\cite{leviathan2023fast}; request admission order~\cite{jiang2026coda}; and prefill microbatching in P/D disaggregated serving~\cite{zhong2024distserve}.
Our characterization (\S\ref{sec:motiv:levers}, \textbf{Findings~1-6}) shows that these levers influence multiple bubble sources and interact with one another.
Tuning one lever to reduce a particular bubble can leave other bubbles dominant or increase them.
For example, changing prefill chunk size alters the mix of computation across stages and can change the best layer partition.
Longer speculative drafts also increase decoding work per round, changing which prefill chunk size best balances prefill and decoding.
This motivates our work on co-optimizing these levers with a unified view of where pipeline capacity is lost.

We present \sys, a pipeline-parallel LLM serving system that co-optimizes these levers through offline configuration and online scheduling.
The two time scales follow the different causes of the bubbles.
Uneven work across stages is a structural property of the model and its placement~\cite{narayanan2019pipedream,wang2024improving}, and changing the pipeline-parallel layer partition requires \headwayreview{computationally} costly reconfiguration~\cite{li2023alpaserve}, so \sys sets the partition offline.
In contrast, the balance of prefill and decoding work, request concurrency, and token acceptance change during serving~\cite{yu2022orca,liu2024optimizing}, so \sys adjusts the remaining levers online.
Offline, \sys selects a layer partition by profiling candidates on a representative workload with the online scheduler enabled, reducing stage bubbles while accounting for the scheduler's online chunk-size and speculative-length choices (\S\ref{sec:design:colocated}).

Online, \sys tailors its scheduling to the serving architecture, because co-located and P/D disaggregated serving differ in the levers they expose and in the bubbles they incur.
In co-located serving (\S\ref{sec:design:colocated}), the online scheduler jointly adapts the prefill chunk size and the speculative length to reduce microbatch and memory bubbles as workload demand changes.
Its decisions balance prefill and decoding progress against execution and memory \headwayreview{computational} costs, with latency guards for TTFT (time-to-first-token) and TPOT (time-per-output-token)~\cite{zhong2024distserve}.
For P/D disaggregated serving (\S\ref{sec:design:pd}), \sys also coordinates request admission and prefill microbatching.
The decode scheduler balances microbatch loads using the selected speculative length and available memory, while the prefill scheduler adjusts microbatch budgets to deliver completed requests to decoding earlier and reduce handoff bubbles.

This paper targets four types of bubbles in pipeline-parallel LLM serving, characterizes how the levers serving engines already expose move them, and co-optimizes these levers offline and online to push the bubbles toward their limit.
Our contributions are summarized as follows:

\noindent\textbf{\underline{Contribution 1:}} We present a detailed characterization of pipeline bubbles that motivates the design of \sys (\S\ref{sec:motivation}).
We distinguish stage, microbatch, memory, and P/D handoff bubbles (\S\ref{sec:motiv:types}) and show in six findings that layer partitioning, prefill chunk size, speculative decoding draft length, request admission, and prefill microbatching each move more than one bubble and interact with one another (\S\ref{sec:motiv:levers}).
These findings show why tuning one lever at a time falls short and call for co-optimizing the levers.

\noindent\textbf{\underline{Contribution 2:}} We design and implement \sys (\S\ref{sec:design}, \S\ref{sec:impl}).
Offline, it selects a pipeline-parallel layer partition by profiling candidates with the online scheduler enabled (\S\ref{sec:design:colocated}).
Online, it jointly schedules prefill chunk size and speculative decoding draft length under TTFT and TPOT guards for co-located serving (\S\ref{sec:design:colocated}), and coordinates request admission and prefill microbatching for P/D disaggregated serving (\S\ref{sec:design:pd}).
We integrate \sys with vLLM in about 14.4\,KLoC, with runtime draft-length adaptation, dynamic memory allocation, and online execution feedback (\S\ref{sec:impl}).

\noindent\textbf{\underline{Contribution 3:}} We evaluate \sys on Qwen3.8-27B and Qwen3.5-122B-A10B with replayed coding-agent sessions in co-located and P/D disaggregated serving (\S\ref{sec:eval}).
Over state-of-the-art pipeline-parallel LLM serving systems, \sys improves co-located token throughput by 17.5\% on average and up to 32.6\% and request throughput by 16.4\% on average and up to 32.4\%, while reducing mean time between output tokens by 17.6\% and mean end-to-end latency by 13.7\% on average with comparable TTFT.
\sys raises P/D token throughput by 19.2\% on average (up to 20.8\%).
Runtime analysis further attributes these \headwayreview{improvements} to smaller stage bubbles and to joint chunk-size and draft-length decisions, with scheduling \headwayreview{computational} overhead below 1.6\% of the step interval (\S\ref{sec:eval:rq2}).


\section{Background}
\label{sec:background}

\noindent \textbf{Pipeline-parallel LLM serving.}
Tensor parallelism shards each layer across GPUs and requires frequent
synchronization over high-bandwidth interconnects~\cite{shoeybi2019megatron,narayanan2021efficient}.
Pipeline parallelism instead assigns contiguous layers to stages and
passes activations between them, reducing communication on
PCIe-connected servers~\cite{huang2019gpipe,narayanan2019pipedream,hwang2026revisiting}.
A $P$-stage pipeline keeps $P$ microbatches in flight, serving each request
once every $P$ steps~\cite{kwon2023efficient,zheng2024sglang}.
Each step is limited by its slowest stage; differences in processing time
across stages and microbatches create pipeline bubbles.
Throughput therefore depends on how evenly work is spread across stages
and microbatches, not only on its total.

\noindent \textbf{Co-located and P/D disaggregated serving.}
Co-located serving combines prefill and decode in shared microbatches,
so prefill work affects decode step times.
P/D disaggregation separates the phases across instances\footnote{An instance is a serving engine that runs one replica of the model
on one or more GPUs.} and transfers KV state between
them~\cite{zhong2024distserve,patel2024splitwise,qin2025mooncake,liu2024cachegen,jiang2025hexgen}.
This removes the interference, but the decode instance can only process
requests that the prefill instance has finished and transferred.

\noindent \textbf{Chunked prefill and speculative decoding in a pipeline.}
Two mechanisms decide what work a microbatch carries.
Chunked prefill processes prompt chunks alongside decode requests:
larger chunks raise prefill throughput and shorten time to first token,
while smaller chunks protect decode latency~\cite{agrawal2023sarathi,agrawal2024taming,hwang2026revisiting}.
Speculative decoding drafts $K$ tokens per request and verifies them in
one target-model pass, producing one to $K{+}1$ valid tokens per
round~\cite{leviathan2023fast,chen2023accelerating,miao2024specinfer}.
Some models integrate multi-token prediction heads~\cite{gloeckle2024better,liu2024deepseek,qwen2025qwen3next}.
Increasing the speculative length $K$\footnote{We use speculative length and $K$
interchangeably for the number of tokens drafted per request
in each round.} raises verification work across
stages and drafting work on the last stage, and requires more
recurrent-state memory in hybrid models~\cite{gu2023mamba,yang2024gated,qwen2025qwen3next}.
Thus, these levers affect both stage and microbatch times, while
speculative length also affects request concurrency through its memory demand.

\noindent \textbf{Systems for pipeline-parallel serving.}
Existing systems optimize pipeline execution~\cite{zhang2025td,he2025sipipe,yin2025specpipe}
or speculative length~\cite{liu2024optimizing,li2025adaserve,vllm2025speculative}.
Sarathi-Serve combines chunked prefill with stall-free
batching~\cite{agrawal2024taming}.
Hwang and Ahn adapt chunk size under TTFT and TPOT objectives and balance
decode requests through delay scheduling~\cite{hwang2026revisiting}.
Each of these systems targets one source of bubble and leaves the
others unexploited. \sys instead starts from the pipeline as a whole.
It characterizes where the pipeline's time goes, across its stages, across
its microbatches, in the memory that bounds them, and in the handoff
from prefill to decode in disaggregated inference, and schedules stage
and microbatch execution against all four, pushing the pipeline's
bubble time toward its limit. The complete background and related-work discussion appear in \autoref{app:extended-background}.

\section{Motivation and Characterization}
\label{sec:motivation}

This section introduces the four types of bubbles that arise in
pipeline-parallel LLM serving and describes what minimizing each of
them requires.

\subsection{Different Types of Bubbles}
\label{sec:motiv:types}

\begin{figure}[t]
\centering
\includegraphics[width=0.85\columnwidth]{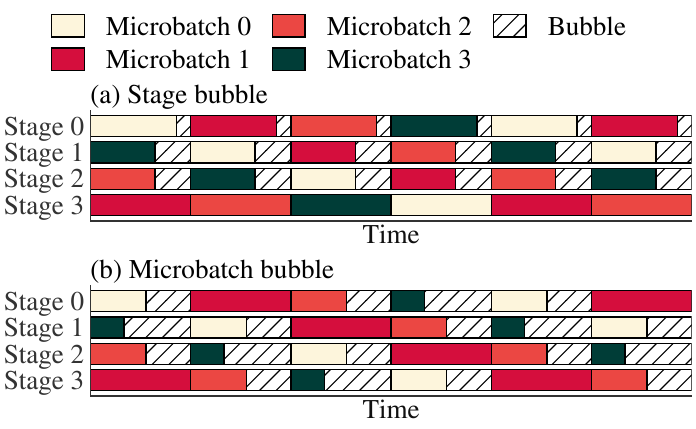}
\vspace{-0.5em}
\caption{Stage and microbatch bubbles of a pipeline on four stages; colors mark microbatches and hatching marks a bubble. (\textbf{\underline{a}})~The stages take different times on the same microbatch, so each microbatch advances at the pace of its slowest stage. (\textbf{\underline{b}})~Consecutive microbatches take different times on the same stage, so the pipeline advances at the pace of its heaviest microbatch.}
\label{fig:bubbles}
\end{figure}

\noindent \textbf{B1: Stage bubble.}
A \stagebubble is the time a stage idles because another stage takes
longer on the same microbatch (Figure~\ref{fig:bubbles}a). It arises
from uneven work across the stages: layers are indivisible, the first
stage carries the embedding, the last stage carries the language-model
head and the drafter, and in a mixture-of-experts model the \headwayreview{computational} cost of a
layer varies with depth. The microbatch advances only when its slowest
stage finishes, so every other stage idles for the difference on every
step.\footnote{A step is a scheduling round with one microbatch per stage.} Removing the stage bubble therefore means
balancing the stages, which decides how much of each step every GPU
computes.

\Needspace{3\baselineskip}\noindent \textbf{B2: Microbatch bubble.}
A \mbbubble is the time a stage idles because the microbatch ahead of it
takes longer than the one it holds (Figure~\ref{fig:bubbles}b). It
arises from uneven work across consecutive microbatches: one carries a
prefill chunk and the next only decodes, the microbatches hold different
numbers of requests, and speculative decoding verifies $K{+}1$ tokens
per decode request, multiplying the work difference that unequal
request counts create. The pipeline advances at the pace of its heaviest microbatch, so
the stage holding a light microbatch idles until the heavy one clears.
Removing the microbatch bubble therefore means equalizing the
microbatches, which decides how much of each step the whole pipeline
computes.

\noindent \textbf{B3: Memory bubble.}
A \membubble is memory capacity tied up in unnecessary reservations
while ready requests await memory. It arises as sessions arrive and finish,
changing concurrency, and contexts grow across turns, increasing memory
demand per request. When memory allocations fail to follow these changes,
unnecessary reservations limit request concurrency. For example,
engines reserve recurrent state for the largest speculative length even
while a shorter one runs (\S\ref{sec:background}). Reducing this
bubble means adapting allocation to changes in workload and
memory demand.

\noindent \textbf{B4: P/D handoff bubble.}
The \pdbubble is specific to disaggregated serving, where prefill and
decode run on separate instances. It is the time the decode instance
spends idle while waiting for work from the prefill instance. It arises
when prefill processing delays the arrival of ready requests. Reducing this
bubble therefore means coordinating prefill and decode to keep the
decode instance busy.

\subsection{Levers on the Bubbles}
\label{sec:motiv:levers}

\noindent \textbf{Characterization setup.}
We evaluate Qwen3.8-27B and Qwen3.5-122B-A10B, hereafter 27B and 122B,
on the PCIe-connected RTX PRO 6000 server described in
\S\ref{sec:eval:setup}, using two and four pipeline stages. We replay multi-turn
coding-agent sessions from OpenSWE~\cite{ahmad2026open}, preserving request order within
each session and enabling prefix reuse across successive turns.
Fixed-window replays run for 300\,s at request concurrency cap
$C{=}64$. Constructed workload comparisons instead pool
valid output tokens over finite-workload completion time
across two runs.

\noindent \textbf{Layer partitioning affects the stage bubble.}
Layer partitioning assigns consecutive groups of model layers to
stages, controlling the computation each stage performs.
We first keep the decode workload fixed and change only the layer
partition. We write a partition as the number of consecutive layers on
each stage in pipeline order, so 32/32 splits the 64 layers of 27B
evenly. With equal layer counts, the final stage is the slowest
for both models. Moving layers
away from it changes the 27B partition from 32/32 to 38/26 and
reduces the stage gap from 9.6 to 2.1\,ms; throughput rises from
540.51 to 625.52 tokens/s, a 15.7\% \headwayreview{improvement}. For 122B with four stages, changing
12/12/12/12 to 12/13/13/10 reduces the gap from 5.5 to 2.5\,ms
and raises throughput from 747.98 to 835.86 tokens/s, an 11.7\%
\headwayreview{improvement}. Partitioning addresses the stage bubble by balancing stage execution times.
We find these partitions iteratively: each iteration moves layers from
the bottleneck stage to an underutilized stage based on live profiling,
then updates the stage times. \sys automates this procedure offline
(\S\ref{sec:design:colocated}).

\finding{1}{Equal layer counts do not imply equal stage times.
Layer partitioning should reduce the \textbf{stage bubble} (B1) by balancing
execution time across stages.}

\noindent \textbf{Prefill chunk sizing affects the microbatch bubble.}
Chunk size ($c$) controls how many prompt tokens a step processes.
We use the 27B model to compare six chunk sizes across five
workloads (W1-W5) constructed from subsampled OpenSWE
prompts, as shown in Table~\ref{tab:char:chunks}. For W1, reducing the chunk from
1024 to 512 tokens raises throughput from 165.78 to 172.54 tokens/s
(4.1\%), while accumulated activation waiting falls from
520.72 to 202.37\,ms. W2 benefits from a smaller chunk:
reducing it from 512 to 256 tokens raises throughput from
355.07 to 370.05 tokens/s (4.2\%), while the corresponding
waiting falls from 337.27 to 31.39\,ms. Across the five workloads, the best measured
chunk size ranges from 128 tokens for W3 and W4 to 384 for W2
and W5 and 512 for W1.
For instance, W1's short decodes finish before prefills at
$c{=}128$, whereas $c{=}512$ advances more prompt work per step
and eliminates this prefill-only tail. Adaptive chunk sizing should therefore
follow the current mix of prefill and decode work to reduce
waiting between successive microbatches.

\begin{table}[ht]
\centering
\caption{Throughput (tokens/s) across prefill chunk sizes ($c$)
and workloads. W1: light decoding; W2: reference; W3: heavy decoding;
W4: light prefill; W5: long-context decoding.
Bold marks the highest throughput across chunk sizes for each workload.}
\label{tab:char:chunks}
\setlength{\tabcolsep}{3pt}
\renewcommand{\arraystretch}{1.0}
\resizebox{\columnwidth}{!}{%
\small\begin{tabular*}{\columnwidth}{@{\extracolsep{\fill}}r | r | r | r | r | r@{}}
\hline
$c$ & W1 & W2 & W3 & W4 & W5 \\
\hline
128 & 101.30 & 361.32 & \textbf{993.39} & \textbf{431.79} & 225.78 \\
\hline
256 & 140.13 & 370.05 & 965.69 & 428.20 & 254.58 \\
\hline
384 & 167.18 & \textbf{374.54} & 990.19 & 429.54 & \textbf{260.35} \\
\hline
512 & \textbf{172.54} & 355.07 & 961.90 & 423.74 & 250.61 \\
\hline
768 & 165.12 & 363.06 & 946.98 & 420.63 & 222.86 \\
\hline
1024 & 165.78 & 353.03 & 943.98 & 421.17 & 249.68 \\
\hline
\end{tabular*}}
\end{table}

\finding{2}{The best chunk size depends on the workload.
Adaptive chunk sizing should reduce the \textbf{microbatch bubble} (B2) by
balancing prefill and decode work.}

\noindent \textbf{Speculative length affects microbatch and memory bubbles.}
Speculative length ($K$) is the number of tokens drafted per
request before each verification round.
We use Qwen3.5-122B-A10B to compare speculative
lengths ($K\in\{3,4,5\}$) across
two subsampled OpenSWE workloads at three concurrency levels,
with a fixed layer partition and prefill chunk size
(Table~\ref{tab:char:spec}). For the long-context workload (W2)
at 48 concurrent requests, increasing $K$ from 3 to 5 raises
throughput from 684.19 to 760.82 tokens/s (11.2\%). At 64
requests, the preference reverses: $K{=}3$ achieves 765.49
tokens/s, exceeding $K{=}5$ by 8.0\%, while observed peak concurrency rises from
48 at $K{=}5$ to 64 at $K{=}3$.
The short-context workload (W1) similarly favors $K{=}4$
at 32 requests but $K{=}3$ at 64, where it outperforms $K{=}5$
by 25.0\%.
For instance, W2 favors $K{=}3$ at high concurrency,
as shorter speculation reserves less state memory and can
accommodate more long-context requests within the same memory budget. Longer speculation can increase token progress per round but
also computation and recurrent-state memory. Speculative length should balance
token \headwayreview{improvements} with compute and memory \headwayreview{requirements}.

\begin{table}[ht]
\centering
\caption{Throughput (tokens/s) by speculative length ($K$) and
request concurrency ($C$) for short (W1) and long (W2)
workloads. Bold marks the highest throughput for each workload and $C$.}
\label{tab:char:spec}
\setlength{\tabcolsep}{3pt}
\renewcommand{\arraystretch}{1.0}
\resizebox{\columnwidth}{!}{%
\small\begin{tabular*}{\columnwidth}{@{\extracolsep{\fill}}r | r | r | r | r | r | r@{}}
\hline
& \multicolumn{3}{c|}{W1} & \multicolumn{3}{c}{W2} \\
\hline
$K$ & $C{=}32$ & $C{=}48$ & $C{=}64$ & $C{=}32$ & $C{=}48$ & $C{=}64$ \\
\hline
3 & 869.20 & 1078.87 & \textbf{1243.17} & 574.56 & 684.19 & \textbf{765.49} \\
\hline
4 & \textbf{879.63} & 1071.28 & 1191.23 & 626.34 & 726.11 & 716.24 \\
\hline
5 & 852.14 & \textbf{1087.15} & 994.63 & \textbf{627.70} & \textbf{760.82} & 708.94 \\
\hline
\end{tabular*}}
\end{table}

\finding{3}{The best speculative length varies by workload.
Adapting this length should reduce
\textbf{microbatch} and \textbf{memory bubbles} (B2, B3) by balancing token \headwayreview{improvements}
with compute and memory \headwayreview{requirements}.}

\noindent \textbf{Layer partitioning and chunk sizing are coupled.}
Chunk sizing changes the mix of prefill and decode computation
in each microbatch, altering the stage balance of a given layer
partition. We report five fixed chunk sizes from 128 to 2048 tokens across four layer
partitions, from 32/32 to 38/26 in steps of two
layers, for 27B on OpenSWE
(Table~\ref{tab:char:joint-partition}). The highest measured
throughput comes from 38/26 at 128 tokens, 36/28 at 256 and
512 tokens, and 34/30 at 1024 and 2048 tokens. With a 512-token
chunk, choosing 36/28 instead of 32/32 raises throughput from
526.66 to 573.65 tokens/s (8.9\%). Partitioning should follow the current chunk size to balance
computation across pipeline stages.

\begin{table}[ht]
\centering
\caption{Throughput (tokens/s) across chunk sizes ($c$) and layer
partitions for 27B on OpenSWE.
Bold marks the highest throughput across partitions for each $c$.}
\label{tab:char:joint-partition}
\setlength{\tabcolsep}{3pt}
\renewcommand{\arraystretch}{1.0}
\resizebox{\columnwidth}{!}{%
\small\begin{tabular*}{\columnwidth}{@{\extracolsep{\fill}}r | r | r | r | r@{}}
\hline
$c$ & 32/32 & 34/30 & 36/28 & 38/26 \\
\hline
128 & 197.49 & 212.38 & 225.15 & \textbf{226.05} \\
\hline
256 & 373.56 & 387.80 & \textbf{407.54} & 403.53 \\
\hline
512 & 526.66 & 555.73 & \textbf{573.65} & 545.84 \\
\hline
1024 & 500.11 & \textbf{510.02} & 502.61 & 486.82 \\
\hline
2048 & 452.40 & \textbf{463.78} & 450.85 & 432.50 \\
\hline
\end{tabular*}}
\end{table}

\noindent \textbf{Chunk sizing should be co-optimized with speculative decoding.}
Speculative length changes decode work per verification round,
so its preferred chunk size can change. We compare 27B
workloads constructed from subsampled OpenSWE prompts with
short (W1) and long (W2) outputs
(Table~\ref{tab:char:joint-chunk-k}).
For W2 at $K{=}4$, reducing the chunk from 768 to 384 tokens
raises throughput from 363.51 to 376.71 tokens/s (3.6\%).
At $K{=}8$, the preference reverses: increasing the chunk from
384 to 768 raises throughput from 379.10 to 387.77 tokens/s
(2.3\%). At $K{=}8$, increasing $c$ from 512 to 768 also
improves W1 from 160.00 to 169.99 tokens/s (6.2\%). All three orderings
hold in both repetitions, and the two repetitions of each of these cells
differ by less than 0.7\%. Chunk sizing should follow speculative length to balance
the progress of prefill and decode requests.

\begin{table}[ht]
\centering
\caption{Throughput (tokens/s) across speculative lengths ($K$) and
prefill chunk sizes ($c$) for 27B OpenSWE workloads.
W1 and W2 denote short- and long-output workloads.
Bold marks the highest throughput across $c$ for each workload and $K$.}
\label{tab:char:joint-chunk-k}
\setlength{\tabcolsep}{3pt}
\renewcommand{\arraystretch}{1.0}
\resizebox{\columnwidth}{!}{%
\small\begin{tabular*}{\columnwidth}{@{\extracolsep{\fill}}r | r | r | r | r | r | r@{}}
\hline
& \multicolumn{3}{c|}{W1} & \multicolumn{3}{c}{W2} \\
\hline
$K$ & $c{=}384$ & $c{=}512$ & $c{=}768$ & $c{=}384$ & $c{=}512$ & $c{=}768$ \\
\hline
2 & \textbf{161.52} & 155.46 & 151.37 & \textbf{320.04} & 316.23 & 303.89 \\
\hline
4 & 168.03 & \textbf{170.20} & 164.15 & \textbf{376.71} & 356.08 & 363.51 \\
\hline
6 & 159.66 & \textbf{167.05} & 161.06 & 373.89 & \textbf{384.29} & 374.39 \\
\hline
8 & 150.02 & 160.00 & \textbf{169.99} & 379.10 & 374.33 & \textbf{387.77} \\
\hline
\end{tabular*}}
\end{table}

\finding{4}{Layer partitioning, chunk sizing, and speculative
length are coupled. Co-optimization should address
\textbf{stage}, \textbf{microbatch}, and \textbf{memory bubbles} (B1-B3).}

\noindent \textbf{Admission control can reduce decode microbatch bubbles.}
Admission control determines when ready requests enter decoding
and which microbatch receives them. Balancing work across
decode microbatches can narrow their execution-time gaps.
On 27B, we use the short- and long-output settings from
Table~\ref{tab:char:joint-chunk-k} for OpenSWE workloads.
We compare first-come-first-served (FCFS) admission with a measured
oracle (Table~\ref{tab:char:admission}). We run FCFS and three strategies
that place each new decode request in the microbatch with fewer
requests, in the one with fewer context tokens, or alternately, and
report the one with the smallest measured decode microbatch gap.
Balancing request counts gives the smallest gap for both workloads.
For instance, balancing decode microbatch execution times in the
long-output workload lowers the measured gap from 6.55 to 0.85\,ms (87.0\%),
which also increases throughput by 5.5\%.
Admission control thus helps balance decode microbatch execution times.

\begin{table}[ht]
\centering
\caption{Decode microbatch execution times and gaps without
(w/o) and with (w/) measured-oracle admission for 27B.
W1 and W2 denote short and long outputs, as in Table~\ref{tab:char:joint-chunk-k};
times are paired fastest/slowest means.
Bold marks the lower gap per workload.}
\label{tab:char:admission}
\setlength{\tabcolsep}{2.5pt}
\renewcommand{\arraystretch}{1.0}
\resizebox{\columnwidth}{!}{%
\small\begin{tabular*}{\columnwidth}{@{\extracolsep{\fill}}l | r | r | r | r@{}}
\hline
 & \multicolumn{2}{c|}{W1} & \multicolumn{2}{c}{W2} \\
\hline
Admission & Times (ms) & Gap (ms) & Times (ms) & Gap (ms) \\
\hline
w/o & 46.27/50.76 & 4.49 & 47.68/54.23 & 6.55 \\
\hline
w/ & 47.17/48.03 & \textbf{0.87} & 48.93/49.78 & \textbf{0.85} \\
\hline
\end{tabular*}}
\end{table}

\finding{5}{Admission control can reduce decode \textbf{microbatch bubbles} (B2).
It should distribute decode work evenly to balance microbatch execution times.}

\begingroup\clubpenalty=10000
\Needspace{3\baselineskip}\noindent \textbf{Prefill microbatching can reduce the P/D handoff bubble.}
Prefill microbatching groups prompt chunks from one or more
requests into small batches for incremental processing.
We compare execution with and without explicit prefill
microbatching on 27B (Table~\ref{tab:char:handoff}).
Handoff idle counts time when
the decode instance is empty, admitted prefills remain unfinished,
and no admitted request is ready, excluding initial fill and final
drain. In W1, microbatching reduces mean handoff idle from
17.16 to 8.11\,s and raises throughput by 52.7\%.
For W2, idle falls from 9.98 to 1.75\,s, with 53.5\% higher
throughput.
For instance, W1's short decodes quickly exhaust ready work,
so small prefill batches supply ready requests sooner and
shorten the handoff gap. Prefill microbatching should therefore deliver ready
requests promptly to reduce decode idle time during the P/D handoff.
\par\endgroup

\begin{table}[ht]
\centering
\caption{Throughput (tokens/s) without (w/o) and with (w/) prefill
microbatching for 27B.
W1-W5 follow the workload definitions in Table~\ref{tab:char:chunks}.
Bold marks the higher throughput for each workload.}
\label{tab:char:handoff}
\setlength{\tabcolsep}{2.5pt}
\renewcommand{\arraystretch}{1.0}
\resizebox{\columnwidth}{!}{%
\small\begin{tabular*}{\columnwidth}{@{\extracolsep{\fill}}l | r | r | r | r | r@{}}
\hline
 & W1 & W2 & W3 & W4 & W5 \\
\hline
w/o & 224.17 & 889.70 & 1559.74 & 1548.45 & 883.97 \\
\hline
w/ & \textbf{342.19} & \textbf{1365.35} & \textbf{1587.76} & \textbf{1561.81} & \textbf{1219.86} \\
\hline
\end{tabular*}}
\end{table}

\finding{6}{Prefill microbatching can reduce the \textbf{P/D handoff
bubble} (B4). It should supply ready requests promptly to keep
the decode instance busy.}

\section{\sys Design}
\label{sec:design}

Guided by our characterization, \sys coordinates different levers to reduce bubbles in pipeline-parallel LLM serving. We first describe the co-located design in \S\ref{sec:design:colocated}, then its
extension to P/D disaggregated serving in \S\ref{sec:design:pd}.


\vspace{-0.5em}
\subsection{Co-located Design}
\label{sec:design:colocated}

\textbf{Design overview.}
In co-located serving, prefill and decoding share the same instance.
To reduce bubbles, \sys coordinates layer partitioning, prefill chunk
size, and speculative length through offline preparation and an online scheduler.
Offline, profiling guides layer partition selection.
With the partition fixed, the online scheduler jointly adapts chunk size
and speculative length to reduce microbatch and memory bubbles as request
demand changes.

\textbf{Offline layer partitioning.}
\textbf{Finding~1} shows that reducing stage bubbles requires balancing stage
execution times, while \textbf{Finding~4} shows that chunk size and speculative
length affect the best layer partition. We therefore profile each
candidate partition by replaying a representative workload\footnote{Workload characteristics may change during serving, but repartitioning is \headwayreview{computationally} costly. Like AlpaServe's periodic model re-placement~\citep{li2023alpaserve}, \sys could revisit the partition using an updated workload subsampled from online serving and change it only when profiling indicates substantial \headwayreview{improvements} that justify reconfiguration.} with the online scheduler enabled.
The profiles guide shifts of contiguous layer boundaries to move work
from bottleneck stages to less-loaded stages, accounting for sampling
and drafting on the final stage.
We evaluate the candidate partitions on the same workload and retain
the feasible one with the highest measured throughput.

\textbf{Online optimization objective.}
\textbf{Finding~2} motivates limiting prefill delays to decoding, while \textbf{Finding~3}
motivates weighing valid speculative tokens against their execution and
memory \headwayreview{requirements}. Since these choices interact (\textbf{Finding~4}), \sys jointly
selects $c$ and $K$ to increase prefill progress and valid decoding
tokens per unit time while reducing bubbles. We rank feasible candidates by a heuristic estimate of prefill and decoding time:\vspace{-0.3em}
\begin{equation}
\scalebox{0.90}{$\displaystyle
J(c,K)=\widehat{T}(c,K)\left(\frac{Q}{p(c,K)}+
\frac{SH}{\widehat{Y}(K)}\right).
$}
\label{eq:online-objective}
\end{equation}\vspace{-0.3em}
\begin{table}[t]
\centering
\caption{Symbols used in the online optimization objective.}
\label{tab:online-objective-symbols}
\begin{minipage}{\columnwidth}
\small
\begin{tabular}{@{}p{0.18\columnwidth}|p{\dimexpr0.82\columnwidth-2\tabcolsep-\arrayrulewidth\relax}@{}}
\hline
\textbf{Symbol} & \textbf{Meaning} \\
\hline
$c$ & Prefill chunk size (tokens) \\
\hline
$K$ & Speculative draft length (tokens) \\
\hline
$\widehat{T}(c,K)$ & Predicted mean time per completed microbatch (seconds) \\
\hline
$Q$ & Remaining prefill tokens requiring computation \\
\hline
$p(c,K)$ & Mean planned prefill tokens per microbatch\footnote{Empty microbatches are excluded from this mean.} \\
\hline
$S$ & Number of pipeline stages \\
\hline
$H$ & Fixed output-token target per request \\
\hline
$\widehat{Y}(K)$ & Mean valid tokens per request per round \\
\hline
$J(c,K)$ & Candidate score (seconds; lower is better) \\
\hline
\end{tabular}
\end{minipage}
\end{table}

\autoref{tab:online-objective-symbols} summarizes the symbols
used in \autoref{eq:online-objective}.
Multiple microbatches execute concurrently, so $\widehat T$ predicts the
average interval between successive microbatch completions.
The ratio $Q/p$ approximates the number of microbatches needed to clear
the remaining prefill work. Each decoding round produces $\widehat Y$ valid tokens per request on average
and takes approximately $S\widehat T$ seconds. Producing $H$ tokens therefore takes $H/\widehat Y$
rounds, or $SH/\widehat Y$ microbatch completion intervals.
Thus, $\widehat TQ/p$ approximates prefill clearance time, whereas
$S\widehat TH/\widehat Y$ approximates the time to advance each decoding
request by $H$ tokens. Minimizing their sum balances progress in both
phases. Only $\widehat T$ and $\widehat Y$ use calibrated models;
\autoref{app:calibration} details the profiling and calibration procedures for both models.

\textbf{Online model updates.}
During serving, \sys refines the calibrated models using runtime feedback.
Eligible microbatch completion intervals update $\widehat T$, while valid
output-token counts update $\widehat Y$ at the executed $K$.
Subsequent decisions use these updated models to follow changes in
execution \headwayreview{computational} cost and token acceptance.

\textbf{Constraint-1: TPOT.}
Reducing bubbles must not come at the \headwayreview{downside} of slow token delivery.
Larger chunks can delay decoding, and longer speculation helps only
if enough drafted tokens are valid. We first filter candidates
by their predicted time per output token (TPOT).
A change in $K$ applies to subsequent drafts, while outstanding drafts
retain their original length. For instance, reducing $K$ from 6 to 3
still requires verification of six-token drafts already in flight.
We therefore constrain predicted TPOT during the transition and in
steady state, when all drafts use the new $K$.
For the current length $K_0$ and TPOT threshold $\tau$, we require\vspace{-0.3em}
\begin{equation}
\resizebox{0.90\columnwidth}{!}{$\displaystyle
\begin{gathered}
\widehat L_{\rm steady}(c,K)=\frac{S\widehat T(c,K)}{\widehat Y(K)},\qquad
\widehat L_{\rm tr}(c,K)=\frac{S\widehat T_{\rm tr}(c,K)}
{\min\{\widehat Y(K),\widehat Y(K_0)\}},\\
\widehat L(c,K)=\max\{\widehat L_{\rm steady},\widehat L_{\rm tr}\}\le\tau.
\end{gathered}
$}
\label{eq:online-tpot}
\end{equation}\vspace{-0.3em}
Here $\widehat T_{\rm tr}$ and $\widehat T$ are predicted using the current
and proposed verification lengths.
The minimum $\min\{\widehat Y(K),\widehat Y(K_0)\}$ uses the smaller
valid-token yield, giving the larger TPOT estimate for a fixed predicted
processing time. Bounding the maximum
$\max\{\widehat L_{\rm steady},\widehat L_{\rm tr}\}$ by $\tau$
requires both the steady-state and transition estimates to satisfy
the TPOT constraint.
Only candidates passing both checks remain eligible.

\textbf{Constraint-2: TTFT.}
Small chunks can protect TPOT yet leave prompts waiting for prefill.
The scheduler therefore adds a time-to-first-token (TTFT) guard:
prefill capacity should handle new arrivals and clear accumulated
work before waiting prompts reach their TTFT target.
We compute prefill capacity and the required prefill processing rate:\vspace{-0.3em}
\begin{equation}
\scalebox{0.90}{$\displaystyle
\begin{aligned}
\rho(c,K)&=\min\!\left\{\frac{p(c,K)}{\widehat T(c,K)},
\frac{p_{\rm tr}(c,K)}{\widehat T_{\rm tr}(c,K)}\right\},\\
r_{\rm req}&=\headwayreview{\gamma(\lambda+r_{\rm catch})},
\end{aligned}
$}
\label{eq:online-ttft}
\end{equation}\vspace{-0.3em}
Here $p_{\rm tr}$ is the mean planned prefill tokens per microbatch
at current verification lengths, $\lambda$ the incoming uncached-token
rate, $r_{\rm catch}$ the extra rate needed to clear older prompts'
remaining prefill work before their TTFT target, \headwayreview{and $\gamma$ a control factor}
(\autoref{app:calibration}).
The ratio $p/\widehat T$ gives prefill tokens per second; $\rho$
takes the smaller steady-state and transition rates.
The sum $\lambda+r_{\rm catch}$ covers new arrivals and accumulated work.
Among TPOT-eligible candidates, we minimize $J$ subject to
$\rho\ge r_{\rm req}$; if none qualifies, we maximize $\rho$.
This promotes prefill progress without relaxing TPOT; TTFT remains
best effort under insufficient capacity.

\textbf{Online scheduling.}
Each decision epoch spans two pipeline cycles ($2S$ scheduling steps). The supported $(c,K)$ grid is small, so \sys uses
exhaustive search at each epoch with little scheduling \headwayreview{computational} overhead (our implementation averages approximately 0.3\,ms of scheduler decision time per scheduling step). It selects a candidate using the objective and Constraints~1-2.
To avoid frequent switching, ordinary $K$ changes require a minimum
switching interval and sufficient improvement over the best eligible
pair retaining the current $K$: lower $J$, or higher $\rho$ under overload.
Algorithm~\ref{alg:online-scheduler} in \autoref{app:calibration}
details the search procedure.


\vspace{-0.5em}
\subsection{P/D Disaggregation Design}
\label{sec:design:pd}

\textbf{Design overview.}
In P/D disaggregated serving, prefill and decoding execute on separate
instances, each with its own scheduler. \sys reuses offline layer
partitioning and adaptive
speculative decoding from \S\ref{sec:design:colocated} and adds two
online mechanisms: prefill microbatching and admission control.
The prefill scheduler uses microbatching to reduce handoff bubbles,
while the decode scheduler uses admission control to reduce microbatch bubbles.

\textbf{Prefill microbatching.}
\textbf{Finding~6} shows that prefill microbatching reduces handoff bubbles by
making completed requests available for decoding earlier.
Before forming the next prefill microbatch, the prefill scheduler adjusts its
total token budget $B$ by dividing the remaining prefill work $Q$
by the prefill pipeline depth $S_P$. This distributes prefill work
across microbatches while adapting their \headwayreview{compute} budget to changes in demand.
Requests become ready for decoding once both prefill and the required cache
transfers complete. \autoref{app:calibration}
details the \headwayreview{compute} budget rule.

\textbf{Admission control.}
\sys employs admission control to reduce decode microbatch bubbles
(\textbf{Finding~5}). For each request $i$ entering decoding, the decode scheduler
selects a microbatch that minimizes the largest resulting load:\vspace{-0.3em}
\begin{equation}
j^\star=\arg\min_j\max_m\left\{L_m+\mathbf{1}[m=j]w_i\right\}.
\label{eq:pd-admission-objective}
\end{equation}\vspace{-0.3em}
Here $i$ identifies the new decoding request, $j$ a candidate microbatch,
and $m$ ranges over all microbatches. The weight $w_i$ represents the
estimated decoding work of request $i$ in one round, while $L_m$ sums
the weights of requests already assigned to microbatch $m$.
The weighting model is calibrated offline and computes $w_i$ from the
current context length and selected $K$ (\autoref{app:calibration}).
The term $\mathbf{1}[m=j]w_i$ contributes $w_i$ to microbatch $j$ and
zero to all other microbatches. Thus, $L_m+\mathbf{1}[m=j]w_i$ is the
total load of microbatch $m$ after assigning request $i$ to $j$.
For each candidate $j$, the inner maximum ($\max_m$) gives the largest
resulting microbatch load. The outer minimization ($\arg\min_j$)
selects the microbatch $j^\star$ with the smallest such maximum.

The decode scheduler implements {\renewcommand{\equationautorefname}{Equation}\autoref{eq:pd-admission-objective}} using a
greedy algorithm, which assigns each new request $i$ to the lightest
microbatch $j^\star$ to balance decode microbatch execution times and
reduce bubbles. After each assignment, it adds the new request's weight
$w_i$ to the selected microbatch's load $L_{j^\star}$, so subsequent
assignments account for the newly added work. A memory guard avoids
head-of-line blocking: if memory blocks the queue head, the scheduler
selects a feasible alternative among a fixed number of subsequent queued
requests using the same objective. A bypass guard prevents starvation:
the scheduler stops moving later requests ahead once the head has been
bypassed a preset number of times or blocked beyond a
preset duration.
Algorithm~\ref{alg:pd-admission} in \autoref{app:calibration} details the greedy procedure.

\textbf{Coordination with adaptive speculative decoding.}
Adaptive speculative decoding changes the work and memory demand of each
decoding request. The decode scheduler therefore recomputes $w_i$ and
$L_m$ using the selected $K$ when making admission decisions. Its memory
guard also accounts for verification work still in flight, so reducing $K$
does not free memory needed to verify earlier drafts.
With both admission control and adaptive speculative decoding, \sys
balances decode microbatch loads and adapts speculative work to reduce
microbatch and memory bubbles.


\vspace{-0.5em}
\section{Implementation}
\label{sec:impl}

\textbf{Runtime integration.}
We implement \sys in approximately 14.4\,KLoC of Python schedulers,
worker extensions, and supporting runtime modules. Integrated with
vLLM~\cite{kwon2023efficient}, scheduler subclasses access request queues, memory availability,
and execution feedback at batch formation and admission. They pass
the resulting batches and speculative lengths to worker extensions
through scheduler outputs. vLLM retains model execution, sampling,
preemption, and the underlying block-pool infrastructure. We reuse
vLLM's LMCache connector~\cite{liu2025lmcache} for cache handoff
between P/D instances.

\textbf{Adaptive speculative execution with dynamic memory allocation.}
\sys supports dynamic $K$ through adaptive recurrent-state memory
allocation and startup-captured CUDA graphs~\cite{gray2019cudagraphs}.
The allocator preserves the maximum-$K$ logical scratch
layout while backing only required slots with physical pages. Before
increasing $K$, the scheduler checks available capacity and allocates
additional pages as needed. When $K$ decreases, it retains
pages needed by outstanding verification and the recurrent state used
by the next step, then detaches unused pages. Completion fences defer their return to the shared pool until
earlier GPU work finishes. With updated block tables, workers reuse
the graph captured for the chosen $K$. Together, these mechanisms let
\sys adapt speculative work and \headwayreview{reliably} reclaim unused state memory
for request admission.

\textbf{Online execution feedback.}
\sys refines the \S\ref{sec:design:colocated} prediction models
with runtime feedback~\cite{gujarati2020clockwork}, linking outputs
to features and verification lengths. Completion callbacks collect elapsed time and valid-token
counts without additional GPU synchronization. To handle changes in
$K$, the scheduler checks executed verification lengths before
accepting a timing window and attributes token-yield feedback to
the executed length. It bounds timing corrections to limit each observation's influence. This feedback keeps scheduling
predictions aligned with changing execution \headwayreview{computational} costs and token acceptance.

\vspace{-1em}
\section{Evaluation}
\label{sec:eval}

We design the evaluation to answer two research questions:
\par\vspace{-0.3em}
\begin{itemize}[leftmargin=*]
\vspace{-0.7em}
  \item \textbf{RQ1:} \emph{How does \textup{\sys} compare to baselines across workloads and model sizes in throughput and tail latency?}
  \par\vspace{-0.4em}
  \item \textbf{RQ2:} \emph{How do \textup{\sys}'s optimizations affect performance, and what runtime behavior explains their benefits?}
\end{itemize}
\vspace{-1.5em}

\subsection{Experimental Setup}
\label{sec:eval:setup}

\noindent \textbf{Hardware environment.}
We use a single server with eight NVIDIA RTX PRO 6000 Blackwell
Server Edition GPUs (96\,GB each) and two AMD EPYC 9555 CPUs
(128 physical cores total).\footnote{\rightskip=0pt plus 1.5em\relax\parfillskip=0pt plus .2\columnwidth\relax\headwayreview{All research reported in this paper, including access to and use of models, datasets, and code, was conducted by the university-affiliated authors.}} The GPUs use PCIe~5.0~$\times16$
links (nominal peak: 64\,GB/s per direction per link).

\noindent \textbf{Models and workloads.}
We evaluate the dense Qwen3.8-27B and MoE Qwen3.5-122B-A10B
models in BF16, using their native MTP heads for speculative
decoding. The 27B model uses two pipeline stages per instance,
and the 122B model uses four. We replay multi-turn coding-agent
sessions from OpenSWE~\cite{ahmad2026open}, preserving recorded
prompts, overlapping prefixes, and sequential requests within
each session. We test fixed and time-varying session concurrency in both
co-located and P/D serving.

\noindent \textbf{Baseline systems.}
We compare \sys with four baselines, two of which we strengthen with
\sys's offline layer partitioning, and with its ablated variants.
(\textbf{\underline{i}})~vLLM: improves serving throughput through efficient
KV-cache management with PagedAttention and continuous
batching~\cite{kwon2023efficient}.
(\textbf{\underline{ii}})~vLLM+\sys-Offline: applies our offline
layer-partition optimization to vLLM to reduce stage bubbles.
(\textbf{\underline{iii}})~H\&A+\sys-Offline: combines our offline layer-partition
optimization with Hwang and Ahn's method~\cite{hwang2026revisiting},
which adjusts prefill work per batch to reduce pipeline bubbles.
(\textbf{\underline{iv}})~vLLM Dynamic: adjusts speculative draft length
according to batch size to balance decoding acceleration and
verification \headwayreview{computational} overhead~\cite{vllm2026dynamic}.
(\textbf{\underline{v}})~\sys Ablations: in P/D serving, we start from vLLM Dynamic and add
offline optimization, admission control, adaptive speculation, and prefill
microbatching one at a time (\S\ref{sec:eval:pd}).

\noindent \textbf{Metrics.}
We measure throughput in valid output tokens per second and
report mean and tail latencies for time to first token (TTFT),
time between emitted tokens (TBT), and end-to-end request
latency (E2E). We also analyze stage and microbatch execution-time
gaps to explain the runtime behavior behind performance differences.

\vspace{-0.5em}
\subsection{End-to-End Co-located Evaluation (RQ1)}
\label{sec:eval:colocated}

\begin{figure}[t]
  \centering
  \includegraphics[width=0.9\columnwidth]{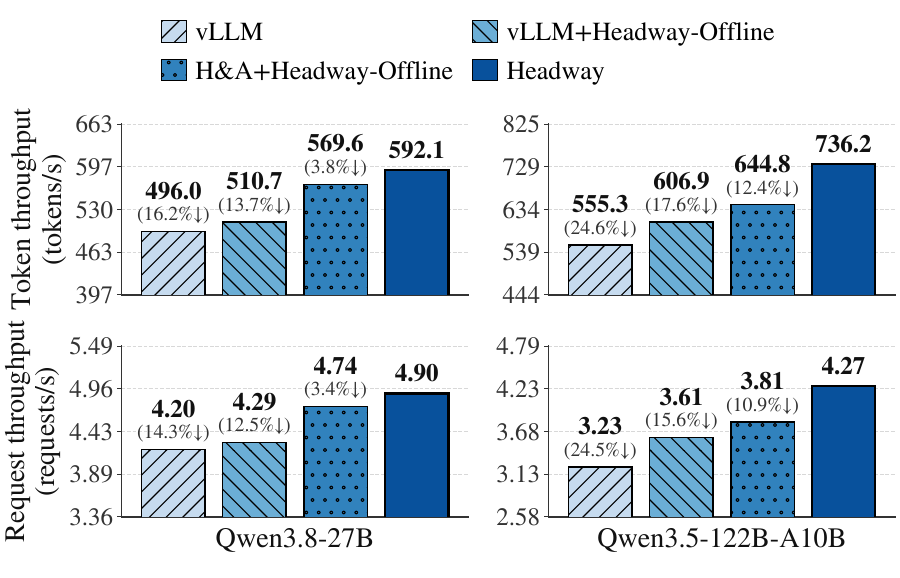}
  \vspace{-0.5em}
  \caption{Co-located token and request throughput. Parentheses show each baseline's percentage decrease from \sys.}
  \label{fig:eval-colocated-throughput}
\end{figure}

\noindent \textbf{End-to-end throughput.}
\sys achieves the highest output-token and completed-request throughput
for both models in Figure~\ref{fig:eval-colocated-throughput}.
Against vLLM, token throughput improves by 19.4\% on 27B and 32.6\%
on 122B, while request throughput increases by 16.7\% and 32.4\%.
Offline layer partitioning alone provides smaller \headwayreview{improvements}: vLLM+\sys-Offline
raises token throughput over vLLM by 3.0\% on 27B and 9.3\% on 122B.
\sys further improves on vLLM+\sys-Offline by 15.9\% and 21.3\%,
showing the additional benefit of online control beyond stage balancing.
H\&A+\sys-Offline is the strongest competing baseline on both models;
\sys exceeds its token throughput by 4.0-14.2\% and request
throughput by 3.5-12.3\%.
The \headwayreview{improvements} in both metrics show that faster token generation also
translates into more completed requests per unit time.
The advantage over H\&A+\sys-Offline highlights the \headwayreview{benefit} of jointly
choosing chunk size and speculative length after selecting a layer partition.
Co-optimizing levers is thus key to reducing different bubbles in pipeline-parallel LLM serving.

\begin{figure}[t]
  \centering
  \includegraphics[width=0.9\columnwidth]{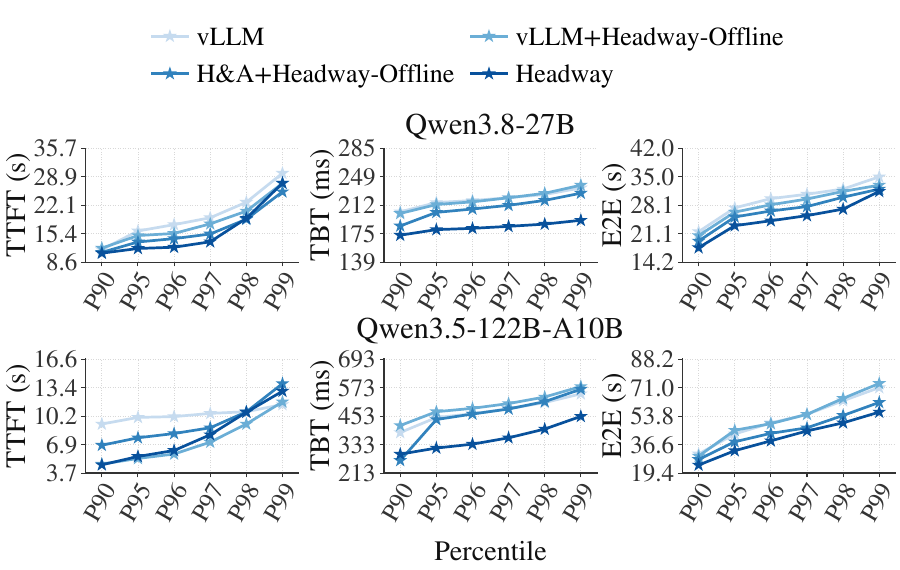}
  \vspace{-0.5em}
  \caption{Co-located TTFT, TBT, and E2E latency at tail percentiles. Lower values indicate better performance.}
  \label{fig:eval-colocated-latency}
\end{figure}

\noindent \textbf{End-to-end latency.}
\sys combines higher throughput with comparable mean TTFT and lower TBT
and E2E latency for both models (Figure~\ref{fig:eval-colocated-latency}).
On 27B, mean TTFT is 8.56\,s, within 4.0\% of H\&A+\sys-Offline's 8.23\,s;
on 122B, it is 3.57\,s, within 6.9\% of vLLM+\sys-Offline's 3.34\,s.
These are the lowest baseline means for each model, and \sys has
lower mean TTFT than the other two baselines in both cases.
Meanwhile, mean TBT decreases by 5.6-23.3\% and mean E2E latency
by 3.8-24.2\% across the two models and three baselines.
The improvements extend to the tail: P99 TBT falls by 15.3-21.8\%
and P99 E2E latency by 1.7-23.5\% across all three baselines.
This pattern is consistent with Constraints~1-2 in
\S\ref{sec:design:colocated}: the TPOT constraint limits delays
introduced by chunk-size and speculative-length choices, while the TTFT
guard maintains sufficient prefill progress for waiting prompts.
Together, these constraints accelerate decoding and request completion
while keeping mean TTFT close to the best baseline.

\begin{figure}[t]
  \centering
  \includegraphics[width=0.9\columnwidth]{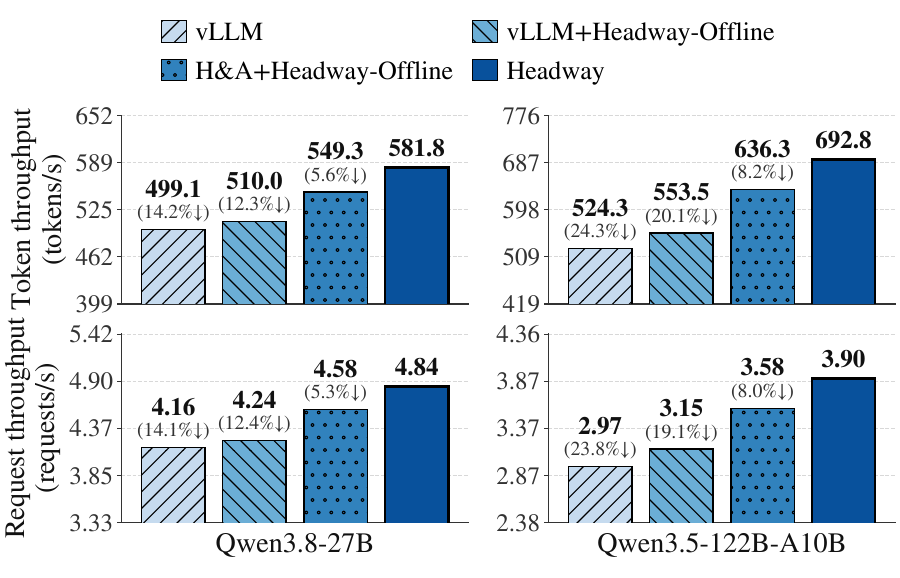}
  \vspace{-0.5em}
  \caption{Token and request throughput under a session-concurrency target
  of $32\rightarrow48\rightarrow64$. Parentheses show each baseline's
  percentage decrease from \sys.}
  \label{fig:eval-colocated-varying}
\end{figure}

\noindent \textbf{Changing workload concurrency.}
We replay OpenSWE with the session-concurrency target increasing from
32 to 48 and then 64, retaining session progress and cache state across
transitions. As shown in Figure~\ref{fig:eval-colocated-varying},
\sys achieves the highest whole-trace token and request throughput
on both models.
Compared with vLLM, token throughput improves by 16.6-32.1\% and
request throughput by 16.4-31.2\%.
H\&A+\sys-Offline remains the strongest baseline, but \sys further improves
its token throughput by 5.9\% on 27B and 8.9\% on 122B, with request-throughput
\headwayreview{improvements} of 5.6\% and 8.7\%.
Coordinating the levers that target pipeline bubbles thus sustains the throughput benefit as demand changes.

\vspace{-0.5em}
\subsection{End-to-End P/D Disaggregation Evaluation (RQ1)}
\label{sec:eval:pd}

\begin{figure}[t]
  \centering
  \includegraphics[width=0.9\columnwidth]{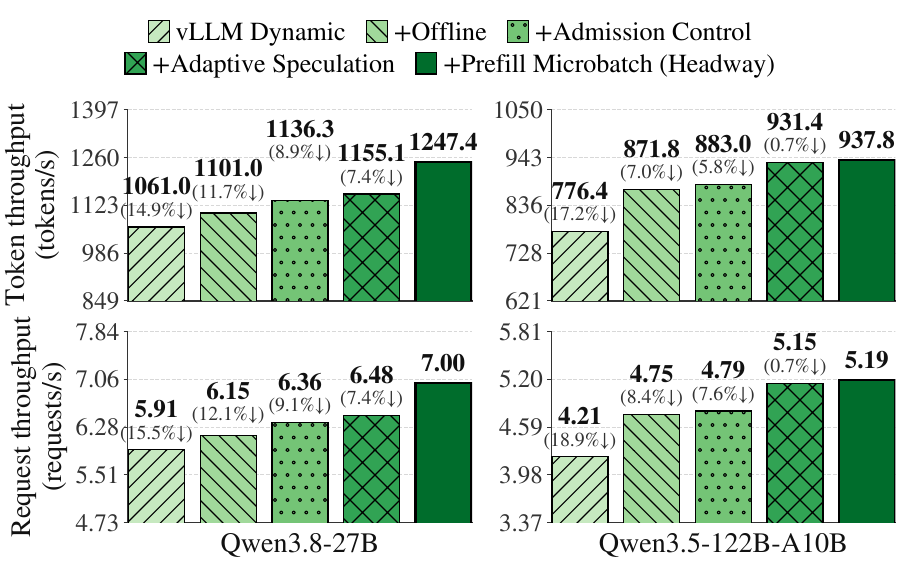}
  \caption{Cumulative P/D throughput ablation. Each configuration adds its optimization to the preceding one. Parentheses show the percentage decrease in throughput from \sys.}
  \label{fig:eval-pd-throughput}
\end{figure}

\noindent \textbf{End-to-end throughput.}
We evaluate P/D-disaggregated serving on OpenSWE with different concurrencies
(Figure~\ref{fig:eval-pd-throughput}).
Starting from vLLM Dynamic, we successively add offline optimization, admission
control, adaptive speculation, and prefill microbatching to obtain the full \sys.
Each configuration retains the preceding optimizations, exposing the incremental
performance \headwayreview{improvement} along this sequence.
Offline optimization increases token throughput by 3.8\% on 27B and 12.3\% on 122B.
Overall, \sys improves token throughput over vLLM Dynamic by 17.6-20.8\%
and request throughput by 18.3-23.3\%.

\noindent \textbf{Admission control and adaptive speculation.}
Adding admission control improves token throughput by 3.2\% on 27B but 1.3\% on 122B.
Its benefit depends on the memory available to balance decode microbatches.
The larger \headwayreview{improvement} on 27B coincides with lower observed decode KV pressure, while
the smaller \headwayreview{improvement} on 122B is consistent with capacity limiting what admission
alone can achieve.
Adaptive speculation contributes more on 122B, adding 5.5\% compared with 1.6\% on 27B.
For the larger model, the trade-off between valid tokens per round and
verification-state memory is more consequential: shortening speculation can
release capacity for more concurrent decoding requests.
The 122B runs combine shorter drafts with higher decode concurrency,
illustrating the value of coordinating speculation with memory-aware admission.

\noindent \textbf{Prefill microbatching.}
The final prefill microbatching step adds 8.0\% token throughput on 27B,
compared with 0.7\% on 122B.
The recorded runs show a larger prefill backlog on 27B for the OpenSWE
workload used here, indicating stronger prefill pressure.
The prefill-side optimization supplies work to the decode instance sooner,
so this optimization has more scope to reduce prefill-handoff bubbles on 27B.
On 122B, the smaller prefill backlog leaves less room for this improvement,
while adaptive speculation addresses the more pronounced memory trade-off.
Together, these results show how adaptive co-optimization addresses different
bubbles across model sizes and workload demands.

\vspace{-0.5em}
\subsection{Optimization Analysis (RQ2)}
\label{sec:eval:rq2}

We examine stage \headwayreview{computational} costs and scheduling decisions to explain
how \sys reduces pipeline bubbles and improves end-to-end performance
in both co-located and P/D-disaggregated serving.

\begin{figure}[t]
  \centering
  \includegraphics[width=\columnwidth]{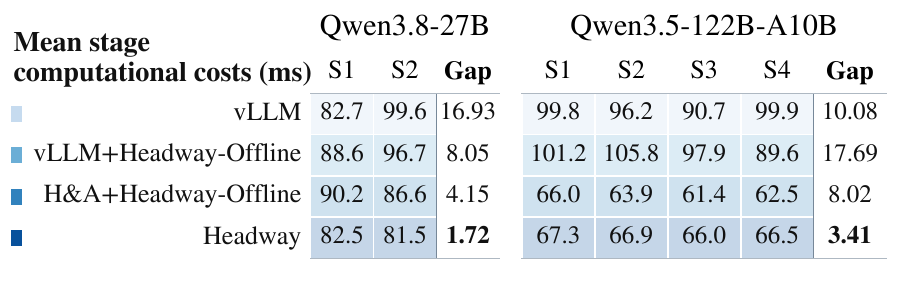}
  \vspace{-1.5em}
  \caption{Mean stage \headwayreview{computational} cost and stage gap per microbatch (ms).
  S1-S4 denote pipeline stages. We average the gap between each microbatch's slowest and fastest stages. Bold marks each model's smallest gap across the four configurations.}
  \label{fig:eval-stage-costs}
\end{figure}

\noindent \textbf{Mean stage \headwayreview{computational} cost per microbatch.}
Figure~\ref{fig:eval-stage-costs} compares the mean stage \headwayreview{computational} cost per microbatch
across the four co-located configurations in Figure~\ref{fig:eval-colocated-throughput}.
Offline partitioning reduces the stage gap from 16.93 to 8.05\,ms on 27B
but increases it from 10.08 to 17.69\,ms on 122B, indicating that partitioning
alone does not ensure balanced work across stages.
Compared with H\&A+\sys-Offline, \sys reduces the stage gap from 4.15 to 1.72\,ms
on 27B and from 8.02 to 3.41\,ms on 122B, reductions of 58.7\% and 57.4\%.
\headwayreview{On 27B, its mean stage computational costs are 82.5/81.5\,ms per microbatch for stages 1 and 2.}
These results highlight the value of coordinating offline layer partitioning
with online adaptive chunk sizing and speculative decoding to reduce the stage bubble.

\begin{table}[ht]
\centering
\caption{122B chunk-size and speculative-length choices under varying concurrency.
HO represents \sys-Offline.
Percentages show how often each value is selected; only the three most
frequent chunk sizes are shown.}
\label{tab:eval:122b-decisions}
\setlength{\tabcolsep}{2pt}
\renewcommand{\arraystretch}{1.0}
\resizebox{\columnwidth}{!}{%
\small\begin{tabular}{@{}l | c | c | c | c | c | c@{}}
\hline
System & \multicolumn{3}{c|}{Chunk Size $c$} & \multicolumn{3}{c}{Spec Length $K$} \\
\hline
H\&A+HO & 128 (23\%) & 3968 (14\%) & 512 (13\%) & \multicolumn{3}{c}{4 (100\%)} \\
\hline
\sys & 1024 (37\%) & 512 (13\%) & 1536 (10\%) & 4 (76\%) & 5 (23\%) & 3 (1\%) \\
\hline
\end{tabular}}
\end{table}

\noindent \textbf{Co-located chunk-size and speculative-length decisions.}
We present a case study of \sys's chunk-size and speculative-length decisions
for the 122B model in co-located serving (Table~\ref{tab:eval:122b-decisions}).
On 122B under varying concurrency, \sys retains $K{=}4$ for 76\%
of choices, extends to $K{=}5$ for 23\%, and shortens to $K{=}3$ for 1\%.
Adaptive speculation thus enables longer drafts selectively, while retaining
shorter lengths to limit verification work and memory demand.
The 122B chunk-size distribution also shifts: \sys most often selects $c{=}1024$
(37\%), and its three most frequent choices, 512, 1024, and 1536, together
account for 60\% of decisions; H\&A+\sys-Offline most often selects $c{=}128$.
This contrast motivates optimizing chunk size alongside $K$:
changing $K$ alters verification \headwayreview{computational} cost and memory reservations, affecting
how much prefill work fits.
Joint scheduling balances prefill and decoding to reduce microbatch and
memory bubbles.

\autoref{app:pd-speculative-decisions} presents the 122B P/D speculative-length case study,
with memory-reclamation evidence in \autoref{app:memory-reclamation}.

\begin{table}[ht]
\centering
\caption{Scheduling time and mean step interval (ms) on 27B.
Overhead is scheduling time divided by the step interval.}
\label{tab:eval:scheduling-overhead}
\setlength{\tabcolsep}{3pt}
\renewcommand{\arraystretch}{1.0}
\resizebox{\columnwidth}{!}{%
\fontsize{8}{9.6}\selectfont\begin{tabular}{l | c | c | c}
\hline
\makebox[\dimexpr(\columnwidth-8\tabcolsep-3\arrayrulewidth)*25/100\relax][l]{\headwayreview{Configuration}} & \makebox[\dimexpr(\columnwidth-8\tabcolsep-3\arrayrulewidth)*25/100\relax][c]{Scheduling} & \makebox[\dimexpr(\columnwidth-8\tabcolsep-3\arrayrulewidth)*25/100\relax][c]{Step interval} & \makebox[\dimexpr(\columnwidth-8\tabcolsep-3\arrayrulewidth)*25/100\relax][c]{Overhead} \\
\hline
Co-located & 0.276 & 82.59 & 0.33\% \\
\hline
P/D decode & 0.732 & 47.47 & 1.54\% \\
\hline
\end{tabular}}
\end{table}

\noindent \textbf{Scheduling \headwayreview{computational} overhead for online scheduling.}
To assess the \headwayreview{computational} cost of online adaptation, we compare \sys's host scheduling
time with the mean scheduling interval on 27B
(Table~\ref{tab:eval:scheduling-overhead}).
The interval is elapsed serving time divided by the step count.
The co-located measurement covers \sys's decision logic, while the P/D
measurement also includes vLLM's batch formation and cache allocation.
Co-located scheduling takes 0.276\,ms per step, or 0.33\% of the
82.59\,ms interval; P/D decode scheduling takes 0.732\,ms, or 1.54\%
of the 47.47\,ms interval.
This supports frequent online adaptation.

\begin{table}[ht]
\centering
\caption{Service-time prediction error on 27B. MAPE is mean absolute
percentage error; lower is better.}
\label{tab:eval:cost-model-accuracy}
\setlength{\tabcolsep}{3pt}
\renewcommand{\arraystretch}{1.0}
\resizebox{\columnwidth}{!}{%
\fontsize{8}{9.6}\selectfont\begin{tabular}{l | c | c | c}
\hline
\makebox[\dimexpr(\columnwidth-8\tabcolsep-3\arrayrulewidth)*25/100\relax][l]{\headwayreview{Configuration}} & \makebox[\dimexpr(\columnwidth-8\tabcolsep-3\arrayrulewidth)*40/100\relax][c]{Timing model} & \makebox[\dimexpr(\columnwidth-8\tabcolsep-3\arrayrulewidth)*20/100\relax][c]{Windows} & \makebox[\dimexpr(\columnwidth-8\tabcolsep-3\arrayrulewidth)*15/100\relax][c]{MAPE} \\
\hline
Co-located & Chunk service time & 1728 & 5.01\% \\
\hline
P/D & Decode service time & 468 & 2.33\% \\
\hline
\end{tabular}}
\end{table}

\noindent \textbf{\headwayreview{Computational cost}-model accuracy for online scheduling.}
To assess the estimates behind these scheduling decisions, we evaluate the
27B service-time models on recorded serving windows
(Table~\ref{tab:eval:cost-model-accuracy}).
The chunk model predicts the mean interval between microbatch completions,
while the decode model predicts the mean decoding-round time per request.
For co-located serving, we use logged predictions computed before feedback
updates. For P/D, we chronologically replay the decode model using observed
concurrency and context length, with feedback only from earlier windows.
The chunk model achieves 5.01\% MAPE, and the P/D decode model achieves 2.33\%.
These low errors support \sys's use of high-fidelity \headwayreview{computational} cost models for online scheduling.

\vspace{-1em}
\section{Conclusion}
\label{sec:conclusion}

\sys reduces pipeline bubbles in pipeline-parallel LLM serving by co-optimizing the levers that serving engines already expose.
Our characterization distinguished different bubbles and showed that the levers each move more than one bubble and interact.
\sys therefore selects the layer partition offline and jointly schedules chunk size and speculative length online, extending to request admission and prefill microbatching in P/D disaggregated serving.
It improves token throughput over state-of-the-art pipeline-parallel LLM serving systems by 17.5\% on average in co-located serving and 19.2\% in P/D disaggregated serving, with lower decoding latency and comparable TTFT.

\bibliography{example_paper}
\bibliographystyle{mlsys2025}

\clearpage

\appendix

\section{Calibration Details}
\label{app:calibration}

\textbf{Chunk service model.}
\label{app:calibration:chunk}
We calibrate each supported chunk setting on the selected layer
partition using representative workloads. The reference service table
uses fresh-prefill observations without decode work or initially cached
prompts. It records the measured times between microbatch completions and
the actual scheduled work, rather than assuming the nominal token
limit is fully used.
The model corrects this reference for query-weighted decode context,
squared prefill query lengths, and prefill-query-prefix products;
reference feature values are subtracted to avoid counting attention
work twice. The prediction target is the host-observed time between successive
nonempty microbatch completions, averaged over a pipeline-depth window.
It is not an individual microbatch's end-to-end latency.
The chunk chooser scans the supported candidates because alignment and
cache boundaries can make productive prefill capacity nonmonotonic.
Its required rate combines arrivals with a catch-up rate based on
waiting tokens and remaining TTFT slack.

\textbf{Speculative service and yield models.}
\label{app:calibration:spec}
We collect fixed-$(c,K)$ calibration arms with the actual verification
width matched to the selected proposal length. Separate nonnegative
regressions model decode-only and mixed/prefill microbatches using query
tokens, decode context work, squared prefill lengths, prefill-prefix
products, and counts of requests, prefills, and unfinished prefills.
Within each arm, the first 75\% of observations train the model and the
last 25\% validate it. Prediction is enabled only for banks or phases
meeting the support requirements, mean absolute percentage error at most
15\%, and 90th-percentile absolute percentage error at most 30\%.
Accepted models are refitted on all calibration rows.
Support checks restrict prediction to the calibrated feature range with
an explicit margin. The timing models are initialized offline and refined
from eligible online completion windows using clipped residuals and
nonnegative coefficients; windows mixing speculative lengths are excluded.
Some configurations freeze the reference model or drafting-\headwayreview{compute} increment.
For each $K$, $\widehat Y(K)$ is the total valid output tokens divided
by the number of eligible request-rounds, pooling calibration observations
across chunk settings. A round contributes accepted draft tokens plus
one correction or bonus token. Online observations extend these totals
at the actually executed $K$, giving a cumulative mean; unobserved
lengths retain their prior values. Prefill, first-output, stale, empty,
and terminal rounds are excluded. The optional $K=6$ extension starts
from an extrapolated yield prior with the weight of 64 rounds, then
uses the same online updates.

\textbf{Parameter acquisition.}
The scheduler selects $c$ and $K$ at decision epochs. It refreshes $Q$
from current request and cache state, computes $p(c,K)$ from each
candidate's scheduling preview, and evaluates $J(c,K)$ from these inputs
and the timing and yield models. The pipeline depth $S$ and token target
$H$ are fixed throughout serving.

\textbf{Coupled previews and \headwayreview{controls}.}
The preview spans one pipeline cycle and respects each request's decode
eligibility. It reserves verification queries before assigning prefill
work through native chunk and cache boundaries. The implementation
realizes chunk sizing through a per-step token limit shared by prefill
and decode; actual prefill work can therefore be smaller than this limit.
For the objective, $Q$ sums the remaining prefill tokens of eligible
requests. With resolved cache state, each contribution is calculated
exactly as the sequence length minus the computed or reusable prefix,
including replay work after preemption. Only waiting requests whose
cache lookup is pending use predicted prefix reuse from earlier requests;
the native lookup replaces this prediction with the actual token count.
The quantity $p(c,K)$ averages planned prefill tokens over nonempty
previewed microbatches; allocation may admit less work than the preview.
The positive integer $H$ specifies how many output tokens per decoding
request are considered when comparing candidates. It is configured
before serving and held fixed across candidates and decision epochs. Larger $H$ weights decoding
progress more strongly relative to prefill clearance. It is not reduced
by emitted tokens or a predicted EOS position; native stopping rules
remove completed requests from subsequent scheduling decisions.
Latency checks cover both already-issued verification widths and the
steady width of the proposed $K$. The guarded variant additionally
preserves prefill demand and prioritizes \headwayreview{acceptable} prefill capacity under
overload. For ordinary length changes, switch epochs and improvement
margins prevent frequent switching; \headwayreview{fallbacks} can override them.
Model-specific refinements can re-evaluate $K$ at the
selected chunk setting; the objective does not imply a global optimum.
Native allocations retain the configured maximum speculative-state
reservation. Changing $K$ alone therefore does not release physical
state memory; cache-pressure guards and native allocation checks
enforce memory limits during scheduling.

\textbf{TTFT demand guard.}
Let $F$ and $D$ be the configured TTFT and TPOT targets.
Among runnable prompts, the aged set contains requests whose age
reaches a configured fraction of $F$. Let $Q_a$ be their remaining
uncached tokens and $\bar a$ their unweighted mean age.
With slack $s=F-\bar a$, the catch-up rate is
\begin{equation*}
r_{\rm catch}=\begin{cases}
0,&Q_a=0,\\
Q_a/s,&Q_a>0\text{ and }s>D,\\
\infty,&\text{otherwise}.
\end{cases}
\end{equation*}
Insufficient slack thus triggers the overload branch, which prioritizes
prefill capacity among TPOT-eligible candidates. Blocked prompts are
excluded from runnable demand. This rate-based guard uses aggregate
waiting age rather than predicting a separate deadline for each request.
\headwayreview{The guarded 122B runs used $\gamma=1$, so the required rate equaled the sum of the arrival and catch-up rates.}

\textbf{Search variants and fallbacks.}
Algorithm~\ref{alg:online-scheduler} describes the demand-aware search
used by the guarded 122B implementation. The primary 27B search uses
the same objective and TPOT screen but omits the prefill-rate filter;
its separate chunk-controller fallback is TTFT-aware. The discrete menu $\mathcal M$ contains candidate $(c,K)$ pairs with
available model estimates. It need not be a Cartesian product: the reference speculative length
can have a wider chunk grid than other lengths.
\textsc{Stabilize} compares the chosen pair $a$ with the best current-$K$
pair $b$ in the same pool $\mathcal P$. An ordinary switch requires an
allowed switching epoch and either $J(a)\le(1-\epsilon)J(b)$ in normal
mode or $\rho(a)\ge(1+\epsilon)\rho(b)$ under overload; otherwise it
retains $b$. If the pool has no current-$K$ pair, it uses the best
reference-$K$ pair when available, bypassing the length-switch interval
for recovery. If neither exists, it accepts $a$ only when switching is
allowed and otherwise invokes the fallback. Normal-mode ties favor the
current $K$, then the smaller chunk; overload ties reverse that order.
The optimized 122B implementation can additionally reconsider $K$ at
the selected chunk when that chunk is outside the primary grid. This
local refinement requires valid previews and positive service estimates,
and rechecks productive prefill, TPOT, and switching \headwayreview{improvement}; it does not
repeat the prefill-rate filter or search over all chunks.
When no pair survives TPOT screening or previews are unavailable, a
separate chunk controller supplies the fallback. Transient fallbacks
can retain $K$, while low-cache recovery restores the reference length.
Fallbacks cannot guarantee per-request TTFT and TPOT targets.

\begin{algorithm}[H]
\caption{Joint search with TTFT and TPOT guards}
\label{alg:online-scheduler}
\begin{algorithmic}[1]
\REQUIRE State $x$, current $K_0$, menu $\mathcal M$, fixed token target $H$, latency targets, \headwayreview{improvement} margin $\epsilon$
\IF{cache is pressured or previews are unavailable}
  \STATE \textbf{return} \textsc{Fallback}$(x,K_0)$
\ENDIF
\STATE Compute $r_{\rm req}$ from TTFT slack and arrivals
\STATE $\mathcal A \gets \emptyset$
\FORALL{$a=(c,K)\in\mathcal M$}
  \STATE Preview current-width and next-$K$ microbatches
  \STATE Predict $J(a)$, $\widehat L(a)$, and $\rho(a)$ using \autoref{eq:online-objective}-\autoref{eq:online-ttft}
  \IF{supported and $\widehat L\le\tau$}
    \IF{$Q=0$ or $\min\{p,p_{\rm tr}\}>0$}
      \STATE $\mathcal A\gets\mathcal A\cup\{a\}$
    \ENDIF
  \ENDIF
\ENDFOR
\IF{$\mathcal A=\emptyset$}
  \STATE \textbf{return} \textsc{Fallback}$(x,K_0)$
\ENDIF
\STATE $\mathcal F\gets\{a\in\mathcal A:Q=0\ \text{or}\ \rho(a)\ge r_{\rm req}\}$
\IF{$\mathcal F\ne\emptyset$}
  \STATE $\mathcal P\gets\mathcal F$; $m\gets\text{normal}$
  \STATE $a\gets\arg\min_{a\in\mathcal P}J(a)$
\ELSE
  \STATE $\mathcal P\gets\mathcal A$; $m\gets\text{overload}$
  \STATE $a\gets\arg\max_{a\in\mathcal P}\rho(a)$
\ENDIF
\STATE $a\gets\textsc{Stabilize}(a,\mathcal P,K_0,m,\epsilon)$
\IF{$a$ is unavailable}
  \STATE \textbf{return} \textsc{Fallback}$(x,K_0)$
\ENDIF
\STATE Apply $c(a)$ and use $K(a)$ for subsequent proposals
\end{algorithmic}
\end{algorithm}

\textbf{P/D prefill microbatch \headwayreview{compute} budget.}
For prefill \headwayreview{compute} budget selection, $Q$ denotes the remaining prefill tokens
requiring computation, as in \S\ref{sec:design:colocated}.
It is calculated from prompt lengths and the actual computed or reusable
prefixes after cache lookup. Requests awaiting cache lookup or transfer
are excluded until ready. Before forming the next prefill microbatch,
the scheduler sets
\begin{equation*}
B=\min\!\left\{B_{\max},\max\!\left\{B_{\min},
u\left\lceil\frac{Q}{S_Pu}\right\rceil\right\}\right\},
\end{equation*}
where $S_P$ is the prefill pipeline depth, $u$ the scheduling unit, and
$B_{\min}\ge u$ and $B_{\max}$ the configured \headwayreview{compute} budget limits.
The native scheduler allocates work within this total token budget and
its cache and memory constraints, so actual microbatches may be smaller.
A configured per-request cap also limits the work assigned to each prompt.

\textbf{P/D admission-weight calibration.}
For each supported $K$, we fit a nonnegative service model using valid
fixed-$K$ profiling windows. Each window records mean decoding
concurrency, total context length, elapsed time, and completed per-request
verification rounds. The timing target is elapsed time multiplied by
mean concurrency and divided by completed verification rounds,
including ordinary admission and cache-transfer \headwayreview{computational} costs. We fit
\begin{equation*}
\widehat T_D=\beta_{0,K}+a_KN+b_K\sum_i\ell_i,
\qquad w_i=a_K+b_K\ell_i,
\end{equation*}
where $N$ is decoding concurrency and $\ell_i$ is the current context
length of request $i$. The coefficients are obtained by nonnegative
least squares, with normalization constants absorbed into their values.
During scheduling, $\ell_i$ is the larger of the prompt length and
computed-token count. Thus, $a_K$ captures a per-request contribution
and $b_K\ell_i$ its context-dependent contribution. The shared intercept
$\beta_{0,K}$ is omitted from individual weights to avoid counting it
repeatedly. These weights support relative microbatch balancing rather
than predicting individual request latencies. Uniform per-$K$ online scaling preserves the weight ordering.

\textbf{Greedy admission procedure.}
Algorithm~\ref{alg:pd-admission} runs at each decoding scheduling step
after the selected $K$ is applied. It first considers whether to move a
memory-feasible request ahead of the previous allocation blocker, then
assigns newly ready decoding requests to microbatches. Candidate $i$ has
score $g_i=\max\{\max_m L_m,\min_m L_m+w_i\}$, the maximum load
after it joins the lightest microbatch ({\renewcommand{\equationautorefname}{Equation}\autoref{eq:pd-admission-objective}}).

\begin{algorithm}[H]
\caption{Greedy decode admission}
\label{alg:pd-admission}
\begin{algorithmic}[1]
\REQUIRE Selected $K$, current time $t$, scan bound $r$, bypass-count limit $b_{\max}$, blocked-time limit $d_{\max}$
\STATE Compute request weights $w_i$ and existing loads $L_m$
\STATE $h\gets$ previous allocation blocker still in a waiting queue
\IF{$h$ exists and still cannot fit in memory}
  \STATE Initialize $(b_h,t_h)\gets(0,t)$ if not yet tracked
  \IF{$b_h<b_{\max}$ and $t-t_h\le d_{\max}$}
    \STATE $\mathcal F\gets$ eligible, memory-feasible requests among the first $r$ entries after $h$
    \IF{$\mathcal F\ne\emptyset$}
      \STATE $i^\star\gets\arg\min_{i\in\mathcal F}g_i$, preferring earlier requests
      \STATE Move $i^\star$ to the front of $h$'s queue
      \STATE $b_h\gets b_h+1$
    \ENDIF
  \ENDIF
\ENDIF
\STATE Continue native scheduling and cache handling
\STATE Record any allocation blocker for the next step
\STATE Recompute $L_m$ for existing decoding requests
\FORALL{requests $i$ starting decoding, in native order}
  \STATE $j^\star\gets\arg\min_m L_m$, ties favor the native group
  \STATE Assign $i$ to microbatch $j^\star$
  \STATE $L_{j^\star}\gets L_{j^\star}+w_i$
  \STATE Clear $i$'s bypass count and blocked timer
\ENDFOR
\end{algorithmic}
\end{algorithm}

The scan counts all queue entries, but only waiting requests with no
computed tokens are eligible. Memory feasibility compares the native
block requirement, including its watermark, with \headwayreview{available} blocks minus
in-flight prefill reservations. Defaults are $r=8$, $b_{\max}=3$, and
$d_{\max}=10$\,s. The timer starts when the request is first observed
blocked, rather than at arrival; each queue promotion consumes one
bypass even if cache handling delays admission. Normal scheduling
continues with or without a bypass. With two stages,
a request joins the lighter microbatch by delaying its first decoding
turn by at most one step. This minimizes
{\renewcommand{\equationautorefname}{Equation}\autoref{eq:pd-admission-objective}} per new request without reassignment.

\clearpage
\section{Memory Reclamation under Adaptive Speculation}
\label{app:memory-reclamation}

We examine a recorded $K=6\rightarrow3$ transition on a 27B P/D decoding
instance configured for up to 128 concurrent requests. This case uses
elastic allocation: unused request-private scratch pages return to the
shared memory pool after outstanding verification completes. The CUDA
memory pool remains allocated. With maximum-length per-request reservations, lowering $K$ does not release these pages.

\begin{table}[ht]
\centering
\caption{Scratch allocation and request concurrency before and after reducing speculative length on 27B.}
\label{tab:memory-reclamation}
\setlength{\tabcolsep}{4pt}
\renewcommand{\arraystretch}{1.0}
\resizebox{\columnwidth}{!}{%
\small\begin{tabular}{l | c | c}
\hline
Metric & $K=6$ & $K=3$ \\
\hline
Scratch pages per running request & 18 & 9 \\
\hline
Mean running concurrency & 69.60 & 107.25 \\
\hline
Mean waiting requests & 57.80 & 20.00 \\
\hline
Mean allocated scratch pages & 1252.80 & 965.25 \\
\hline
\end{tabular}}
\end{table}

Table~\ref{tab:memory-reclamation} reports sample means over the final
five seconds of the respective 15-second periods. During the transition,
the allocator recycled 516 scratch pages after completion fences. Of
these, 507 pages belonged to 58 requests that subsequently continued
generating, demonstrating that reclamation occurred before those requests
finished. Meanwhile, mean running concurrency increased from 69.60 to
107.25, while the waiting queue decreased from 57.80 to 20.00. The lower
scratch footprint therefore accompanied higher decoding concurrency
without enlarging the memory pool.

\clearpage
\section{Extended Background}
\label{app:extended-background}
\nocite{zhang2025lmdeploy,jiang2025demystifying,jiang2025thunderserve,jiang2026boute,he2025efficient}

\noindent \textbf{Pipeline-parallel LLM serving.}
Tensor parallelism shards every layer across GPUs and synchronizes them
at each layer, so it depends on high-bandwidth interconnects such as
NVLink~\cite{shoeybi2019megatron,narayanan2021efficient}. Pipeline
parallelism instead assigns contiguous layers to each GPU, called a
stage, and passes only activations between stages, which suits
PCIe-connected servers and accelerators without
NVLink~\cite{huang2019gpipe,narayanan2019pipedream,hwang2026revisiting,miao2022galvatron}.
A serving engine keeps $P$ microbatches in flight on a $P$-stage
pipeline: at every scheduler step each stage executes one microbatch and
forwards it to the next, so a request is served once every $P$
steps~\cite{kwon2023efficient,zheng2024sglang}. Each step lasts as long
as the slowest stage visit among the microbatches in flight, and every
other stage waits for the remainder. These waits are pipeline bubbles.
They arise when the stages of one microbatch take different times, and
when consecutive microbatches take different times on the same stage.
Throughput therefore depends on how evenly work is spread across stages
and microbatches, not only on its total.

\noindent \textbf{Co-located and P/D disaggregated serving.}
Co-located serving runs prefill and decode on the same
instances, so prefill chunks and decode requests share
microbatches and prefill work changes decode step times.
P/D disaggregation instead places the two phases on separate
instances\footnote{An instance is a serving engine that runs one replica of the model
on one or more GPUs.} connected through a KV
store~\cite{zhong2024distserve,patel2024splitwise,qin2025mooncake,liu2024cachegen}.
This removes the interference, but the decode instance can only process
requests that the prefill instance has finished and transferred.

\noindent \textbf{Chunked prefill and speculative decoding in a pipeline.}
Two mechanisms decide what work a microbatch carries. Chunked prefill
splits a prompt into chunks and schedules one chunk alongside the decode
requests of the same step~\cite{agrawal2023sarathi,agrawal2024taming}.
The chunk size sets the prefill share of each microbatch: larger chunks
raise prefill throughput and shorten the time to first token, while
smaller chunks shorten each step and protect decode
latency~\cite{hwang2026revisiting}. Speculative decoding adds a drafter
that proposes $K$ tokens per request and lets the target model verify
them in one forward pass, producing between one and $K{+}1$ valid tokens per
round~\cite{leviathan2023fast,chen2023accelerating,miao2024specinfer};
multi-token prediction (MTP) heads make the drafter part of the
model~\cite{gloeckle2024better,liu2024deepseek,qwen2025qwen3next}. The
speculative length $K$\footnote{We use speculative length and $K$
interchangeably for the number of tokens drafted per request
in each round.} sets the decode work of each microbatch: every
decode request contributes $K{+}1$ query tokens to verification, and the
$K$ draft steps execute only on the last stage, which holds the
language-model head and the MTP layer. In hybrid models that interleave
gated DeltaNet or Mamba layers with attention, $K$ also sets memory,
because each draft position needs its own copy of the recurrent
state~\cite{gu2023mamba,yang2024gated,qwen2025qwen3next}. Serving
engines pick $K$ from a static table indexed by batch size and reserve
state for the largest $K$ in the
table~\cite{vllm2025speculative,liu2024optimizing}. Consequently, each
of the three levers changes the time of more than one stage and more
than one microbatch, and the speculative length also changes how many
requests fit in memory.

\noindent \textbf{Systems for pipeline-parallel serving.}
Pipeline bubbles in serving have been attacked one mechanism at a time.
Sarathi-Serve limits the latency impact of prefill on decode with
chunked prefill and stall-free batching~\cite{agrawal2024taming}. Hwang
and Ahn adapt the chunk size to the request rate under TTFT and TPOT
objectives and rebalance decode requests across microbatches with delay
scheduling~\cite{hwang2026revisiting}. TD-Pipe alternates a pipeline
between prefill-only and decode-only phases~\cite{zhang2025td}, SiPipe
offloads sampling and transmission to the CPU~\cite{he2025sipipe}, and
SpecPipe fills a pipeline with the speculative tokens of a single
request~\cite{yin2025specpipe}. SmartSpec and AdaServe adapt the
speculation length per request for goodput and service-level
objectives, and vLLM lowers it as the batch
grows~\cite{liu2024optimizing,li2025adaserve,vllm2025speculative}.
Each of these systems targets one source of bubble and leaves the
others unexploited. \sys instead starts from the pipeline as a whole.
It characterizes where the pipeline's time goes, across its stages, across
its microbatches, in the memory that bounds them, and in the handoff
from prefill to decode in disaggregated inference, and schedules stage
and microbatch execution against all four, pushing the pipeline's
bubble time toward its limit.

\clearpage
\section{P/D Speculative-Length Decisions}
\label{app:pd-speculative-decisions}

\begin{table}[ht]
\centering
\caption{122B P/D speculative-length choices. Percentages show how often
each length is used across all decoding rounds.}
\label{tab:eval:pd122-spec-decisions}
\setlength{\tabcolsep}{2pt}
\renewcommand{\arraystretch}{1.0}
\resizebox{\columnwidth}{!}{%
\fontsize{8}{9.6}\selectfont\begin{tabular}{@{}l@{}|@{}c@{}}
\hline
\makebox[\dimexpr0.30\columnwidth-\arrayrulewidth\relax][l]{System} & \makebox[0.70\columnwidth][c]{Spec Length $K$} \\
\hline
\makebox[\dimexpr0.30\columnwidth-\arrayrulewidth\relax][l]{vLLM Dynamic} & \begin{tabular}{c | c | c}\makebox[\dimexpr(0.70\columnwidth-6\tabcolsep-2\arrayrulewidth)/3\relax][c]{5 (14\%)} & \makebox[\dimexpr(0.70\columnwidth-6\tabcolsep-2\arrayrulewidth)/3\relax][c]{6 (42\%)} & \makebox[\dimexpr(0.70\columnwidth-6\tabcolsep-2\arrayrulewidth)/3\relax][c]{7 (44\%)}\end{tabular} \\
\hline
\makebox[\dimexpr0.30\columnwidth-\arrayrulewidth\relax][l]{\sys} & \begin{tabular}{c | c | c | c}\makebox[\dimexpr(0.70\columnwidth-8\tabcolsep-3\arrayrulewidth)/4\relax][c]{3 (6\%)} & \makebox[\dimexpr(0.70\columnwidth-8\tabcolsep-3\arrayrulewidth)/4\relax][c]{4 (3\%)} & \makebox[\dimexpr(0.70\columnwidth-8\tabcolsep-3\arrayrulewidth)/4\relax][c]{5 (34\%)} & \makebox[\dimexpr(0.70\columnwidth-8\tabcolsep-3\arrayrulewidth)/4\relax][c]{6 (57\%)}\end{tabular} \\
\hline
\end{tabular}}
\end{table}

We also present a case study of \sys's speculative-length decisions for the
122B model under P/D disaggregation (Table~\ref{tab:eval:pd122-spec-decisions}).
\sys uses $K{\leq}5$ in 43\% of rounds and $K{=}6$ in 57\%,
whereas vLLM Dynamic uses $K{=}7$ in 44\%.
vLLM Dynamic selects $K$ from a fixed batch-size lookup; \sys instead
balances valid-token yield against verification \headwayreview{computational} cost and the memory available
for admission. \sys therefore co-optimizes speculative decoding and admission
to minimize decode microbatch and memory bubbles.
\autoref{app:memory-reclamation} examines scratch-page reclamation and the accompanying change in decoding concurrency after a live switch to shorter drafts.

\end{document}